\documentclass[sigconf]{acmart}

\AtBeginDocument{%
  \providecommand\BibTeX{{%
    \normalfont B\kern-0.5em{\scshape i\kern-0.25em b}\kern-0.8em\TeX}}}

\copyrightyear{2024}
\acmYear{2024}
\setcopyright{rightsretained}
\acmConference[In2Writing '24]{The Third Workshop on Intelligent and Interactive Writing Assistants}{May 11, 2024}{Honolulu, HI, USA}
\acmBooktitle{The Third Workshop on Intelligent and Interactive Writing Assistants (In2Writing '24), May 11, 2024, Honolulu, HI, USA}
\acmDOI{10.1145/3690712.3690720}
\acmISBN{979-8-4007-1031-5/24/05}

\usepackage{multirow}
\usepackage{array}
\usepackage{vcell}
\usepackage{enumitem}
\usepackage{soul}
\usepackage{subfigure}
\usepackage{float}
\usepackage{cuted}

\newcommand{\out}[1]{{}} 

\begin{document}

\title{Work-in-Progress: An empirical study to understand how students use ChatGPT for writing essays and how it affects their ownership }

\author{Andrew Jelson}
\email{jelson9854@vt.edu}
\affiliation{%
  \institution{Virginia Tech}
  \city{Blacksburg}
  \state{Virginia}
  \country{USA}
  \postcode{24061}
}

\author{Sang Won Lee}
\email{sangwonlee@vt.edu}
\affiliation{%
  \institution{Virginia Tech}
  \city{Blacksburg}
  \state{Virginia}
  \country{USA}
  \postcode{24061}
}

\renewcommand{\shortauthors}{Jelson, Lee}

\begin{abstract}
This paper was a Workshop Paper. See the full paper which will be presented at CHI 2026
:

https://arxiv.org/abs/2501.10551 

\noindent
As large language models (LLMs) become more powerful and ubiquitous, systems like ChatGPT are increasingly used by students to help them with writing tasks.
To better understand how these tools are used, we investigate how students might use an LLM for essay writing, for example, to study the queries asked to ChatGPT and the responses that ChatGPT gives.
To that end, we plan to conduct a user study that will record the user writing process and present them with the opportunity to use ChatGPT as an AI assistant. This study's findings will help us understand how these tools are used and how practitioners --- such as educators and essay readers --- should consider writing education and evaluation based on essay writing.

\end{abstract}

\keywords{}

\maketitle

\section{Introduction}

Over the past few years, HCI researchers have been looking into how we can use Large Language Models(LLMs) as tools to enhance the creation process ~\cite{gero_social_2023,zhang_visar_2023, han_recipe_2023}. One side of the research examines how AI has become relevant to writers~\cite{gero_social_2023, zhang_visar_2023}. For example, ChatGPT can provide assistance in creating ideas or proofreading an essay. Although there is concern about how students might use these LLMs to do the work for them, many instructors view the increasing use of these tools as inevitable and believe that students can still learn effectively through the thoughtful use of AI coding assistants ~\cite{wang2023exploring}. Writing assistant tools have emerged in different ways over the last two decades, and there have been numerous papers looking into their effectiveness on essay writing for students, especially for ESL writers~\cite{huang_effectiveness_2020, dong_using_2021,jayavalan_effectiveness_2018, karyuatry_grammarly_2018, oneill_stop_2019, koltovskaia_student_2020}. While most students find increased performance with Grammarly, for example, they do not effectively use the tool, and only make moderate changes to their drafts. ChatGPT has also been investigated as a tool to help students and researchers in the creation process for natural and computer languages \cite{shoufan_can_2023, liu_check_2023, murillo_engineering_nodate, stark_can_nodate}. As LLMs serve as practical support tools and can produce work of comparable quality to that of humans, it can be a challenge for readers to consider these tools when they consume written content 

One particular domain that shares such challenges is writing education. As these systems become stronger and more efficient, students might use ChatGPT to write their papers. This presents new, unique challenges for education. Several articles have been published in recent years on the creation of policies and expectations of LLM assistance in education \cite{adams_artificial_2022, cotton_chatting_2023, halaweh_chatgpt_2023, biswas_role_2023}, and most of them have come to the conclusion that while the use of LLM will become more prevalent in the future, instructors should prepare ways for students to use it effectively. Other investigators have investigated the different risks and benefits of using these LLMs in educational environments~\cite{anders_is_2023,sok_chatgpt_2023, joyner_chatgpt_2023, Warner_chatgpt_2023,mosaiyebzadeh_exploring_2023}. These papers discuss the different ways in which we can implement policies to help use LLM assistants. 
Other researchers have looked at the effectiveness of ChatGPT in problem-solving process \cite{shoufan_can_2023, rudolph_chatgpt_2023, ali_impact_2023, hilliger_assessing_2022}. These papers have all found that while ChatGPT is effective at replicating human work. 
However, we do not understand how ChatGPT will impact how students learn essential academic skills such as writing. To understand the impact, we first need to understand how writers may use ChatGPT in their writing practice in an educational context.  

Our user study seeks to answer the following research questions. 
\begin{itemize}
    \item RQ1: In what ways do students use LLM-powered tools, in our case ChatGPT, in essay writing?
    \item RQ2: How does using LLM-powered tools affect students' perceived ownership?
\end{itemize}
Understanding the answers to these questions will help us refine how we should design writing assistant tools that integrate LLM-powered intelligence into the tools. Furthermore, this research can be leveraged to grasp the level of trust students place in LLMs, and the various approaches --- policy and learning activities --- instructors can adopt to prepare their students to learn writing skills and utilize these tools effectively.

\begin{figure*}[t]
    \centering
    \includegraphics[width=.9\textwidth]{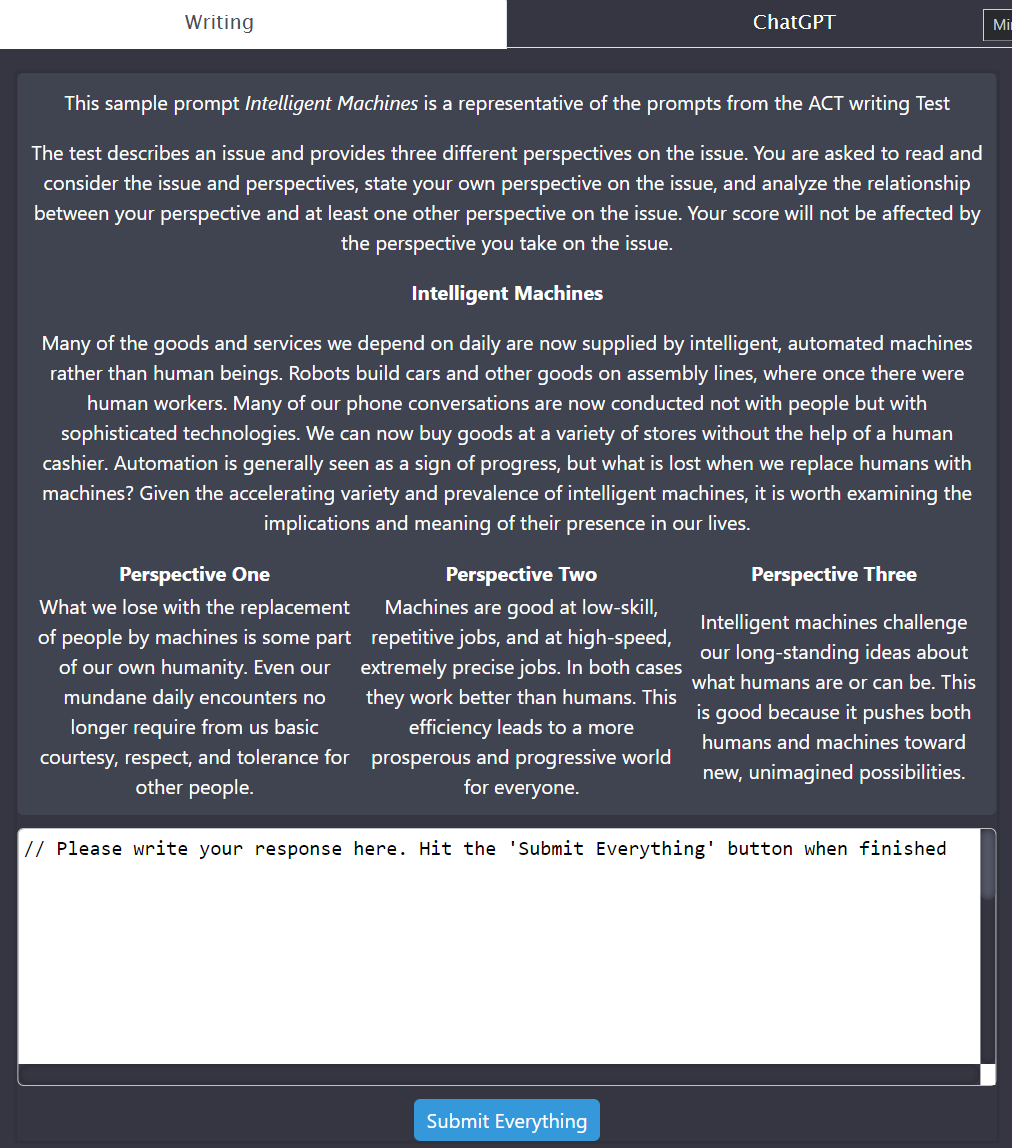}
    \caption{The Essay Recording Page. We used a sample ACT question as the prompt}
    \label{fig:writ_page}
\end{figure*}

\section{Methods}

To understand the different ways students use ChatGPT, we need to track the queries they make and the responses that ChatGPT provides. Since ChatGPT is an independent app, we built a system that makes ChatGPT available within the writing platform so that we can record user interactions on a large scale. Using the tool, we plan to conduct a study asking students to write an essay with ChatGPT assistance and collect data --- query, response, and writing --- for further analysis. We outline the details of the study below.

\begin{figure*}[t]
    \centering
    \includegraphics[width=.9\linewidth]{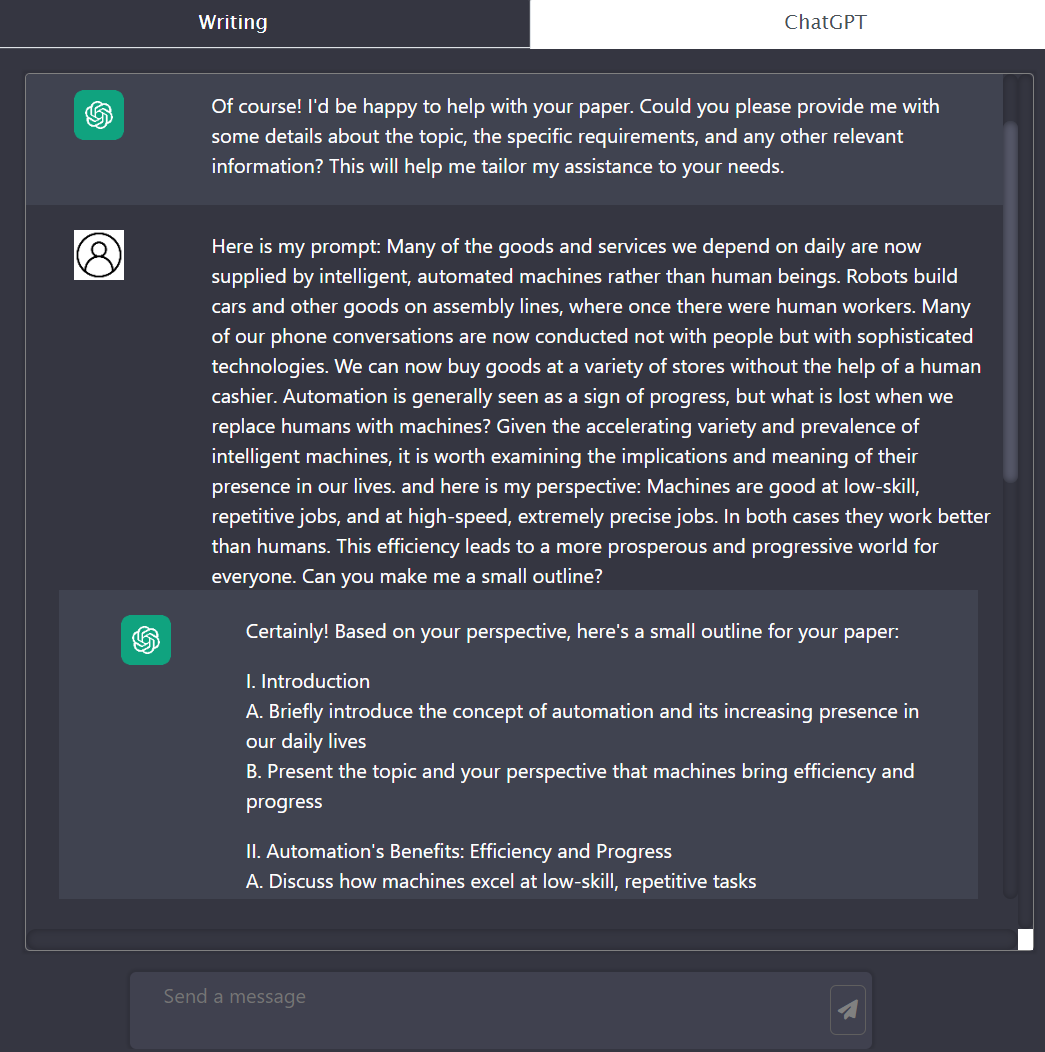}
    \caption{The simulated ChatGPT}
    \label{fig:chat_page}
\end{figure*}

\subsection{Instrument Development}
    Our application has two main features: the ChatGPT feature and a text editor. Both features are developed to track how the essay was written and what kinds of queries and responses are generated in ChatGPT, and we chose a web application to replicate ChatGPT. We want to simulate the experience of using ChatGPT in our design, so we use a tab feature to simulate a modern browser.
    
    The first 'tab' (Fig~\ref{fig:writ_page}) of our application is a writing platform that can record the essay writing process at keystroke levels. The participant will be asked to answer an essay question in a text box that records inputs from the user, tracking cursor position, insertions, deletions, cuts, and pastes performed. We also record the timestamp so that we can see when the user made each edit. Then, we will take this recording to observe and analyze their writing process asynchronously. Having the time stamps allows us to see how they alternate between the editor and the in-house ChatGPT and how they integrated ChatGPT responses into their writing (e.g., pasted text).  Later, we plan on using the recorded scenarios for an educator to evaluate what level of ChatGPT reliance is considered cheating or plagiarism. 
    All of this will be sent to a database on submission. These features were implemented using CodeMirror 5 API and the CodeMirror-Record \cite{Jisuanke2023} files. 

    To track how users use LLM chat tools like ChatGPT, we implement a chat-bot with the Open AI API (model GPT3.5-turbo), shown in Fig~\ref{fig:chat_page}. Users are allowed to ask any question, but the bot will be pre-prompted to be an essay assistance tool. As users use the tool, we record their query and time stamp to see how and when ChatGPT was prompted for assistance during the creation process. This will also be sent to the database on submission.

\subsection{Study Details}
    We will ask the user to write the response to a pre-selected essay question that we have chosen and received from a professor at our university who teaches a junior-level ethics course, i.e.,  Professionalism in Computing. Using this prompt, we will see how university students might use ChatGPT for help with assignments.
    
    Lastly, we will ask the user to complete an exit survey to see how users felt about using ChatGPT when writing their essays. These questions will be focused on how the tool impacts their writing performance and their perceived ownership of the essay (shown in Appendix A.1). As mentioned above, we want to simulate browser tabs for a few reasons. We want the writers to feel that ChatGPT is available to them but not forced upon them. They can choose whether or not to use it, and it is not always on the screen when writing as a distraction.

    For recruitment, we will collect our data from students in an ethics and professionalism course at our university. Most of these students are in their third year and are familiar with essay writing. We will give them a monetary incentive to help with the recruitment process.

\subsection{Data Analysis}

    We plan to analyze our data in multiple different ways. First, we will look directly at the queries made to ChatGPT. Using open coding, we will categorize these questions to see how users prompt the LLM. This will provide us with a better understanding of how people use ChatGPT in the essay creation process.

    Next, we look at the essay itself. As previously stated, the recording features track the user's inputs and store them in our database with timestamps. With this data, we can understand how the response they received from ChatGPT contributes to the writing process by comparing the responses that they get and what new content is added or how the essay is revised immediately after the responses from ChatGPT. This provides us with insight into how the users use LLM-powered tools and their effectiveness. The example metrics that we planned to use include the number of words copied ChatGPT and eventually contributed to the final text, the pace of writing, and the types of generation (e.g., keystroke, pasted text, deleted text, cut-and-pasted text) and how that correlates to ChatGPT usage. 
    
    We also plan to look at where users made changes to their code. As ChatGPT is available throughout the creation process, understanding where the changes in the writing were made is important, for example, if users backtrack and use ChatGPT as a proofreader. This could cause changes to the first line of their writing at the end of the submission, and we will be able to visualize it. Other users could potentially ignore the ChatGPT response and continue with their writing task, so looking at the location will also increase our understanding of how users utilize the LLM in their writing. 
    

\section{Expected Contributions}

    Overall, we expect to gain insight into how users use ChatGPT when writing essays. We will be able to identify patterns in the questions asked and to what extent they implement ChatGPT in their writing. This will be beneficial to instructors who intend to better understand how their students might use ChatGPT and allow them to recognize the level at which their students might use an LLM in their assignment or how to better integrate ChatGPT into their course. This will also benefit software engineers or individuals who wish to create an LLM-powered writing assistant because they can see what features and questions are most common, leading to the development of a better tool.


\bibliographystyle{ACM-Reference-Format}
\bibliography{bibliography}

\appendix
\section{Appendix}
\subsection{Exit Survey}
    Thank you for participating in our study. Please answer the following questions as part of our exit survey.
    
    For the following questions, please answer based on your perceived ownership of the essay: 
    
    \begin{enumerate}
        \item I feel that this is my essay
        \begin{enumerate}
            \item strongly agree, agree, somewhat agree, neutral, somewhat disagree, disagree, strongly disagree
        \end{enumerate}
        \item I feel that this essay belongs to me
        \begin{enumerate}
            \item strongly agree, agree, somewhat agree, neutral, somewhat disagree, disagree, strongly disagree
        \end{enumerate}
        \item I feel a high degree of ownership towards this essay
        \begin{enumerate}
            \item strongly agree, agree, somewhat agree, neutral, somewhat disagree, disagree, strongly disagree
        \end{enumerate}
        \item I feel the need to protect my ideas from being used by others.
        \begin{enumerate}
            \item strongly agree, agree, somewhat agree, neutral, somewhat disagree, disagree, strongly disagree
        \end{enumerate}
        \item I feel that this essays success is my success
        \begin{enumerate}
            \item strongly agree, agree, somewhat agree, neutral, somewhat disagree, disagree, strongly disagree
        \end{enumerate}
        \item I feel this essay was written by me
        \begin{enumerate}
            \item strongly agree, agree, somewhat agree, neutral, somewhat disagree, disagree, strongly disagree
        \end{enumerate}
        \item I feel the need to protect the ideas written in the essay
        \begin{enumerate}
            \item strongly agree, agree, somewhat agree, neutral, somewhat disagree, disagree, strongly disagree
        \end{enumerate}
        \item I do not feel like anyone else wrote this essay.
        \begin{enumerate}
            \item strongly agree, agree, somewhat agree, neutral, somewhat disagree, disagree, strongly disagree
        \end{enumerate}
    \end{enumerate}

    For the following questions, please answer based on your usage of ChatGPT:
    
    \begin{enumerate}
        \item I feel like ChatGPT helped me in the creation process of my writing
        \begin{enumerate}
            \item strongly agree, agree, somewhat agree, neutral, somewhat disagree, disagree, strongly disagree
        \end{enumerate}
        \item I feel like ChatGPT helped me with proofreading my essay
        \begin{enumerate}
            \item strongly agree, agree, somewhat agree, neutral, somewhat disagree, disagree, strongly disagree
        \end{enumerate}
        \item I feel like ChatGPT made my essay better
        \begin{enumerate}
            \item strongly agree, agree, somewhat agree, neutral, somewhat disagree, disagree, strongly disagree
        \end{enumerate}
        \item I liked using ChatGPT as an assistant during my essay writing
        \begin{enumerate}
            \item strongly agree, agree, somewhat agree, neutral, somewhat disagree, disagree, strongly disagree
        \end{enumerate}
        \item My writing would have been better without ChatGPT assistance
        \begin{enumerate}
            \item strongly agree, agree, somewhat agree, neutral, somewhat disagree, disagree, strongly disagree
        \end{enumerate}
    \end{enumerate}
    
    Thank you for completing our survey. Winners of the essay writing competition will receive an email after the study is complete.
    

\end{document}